\documentclass[twocolumn,twocolumcon]{IEEEtran}
\usepackage[T1]{fontenc}
\usepackage{color}
\usepackage{float}
\usepackage{mathrsfs}
\usepackage{amsmath}
\usepackage{amssymb}
\usepackage{graphicx}
\usepackage[unicode=true,
 bookmarks=true,bookmarksnumbered=true,bookmarksopen=true,bookmarksopenlevel=1,
 breaklinks=false,pdfborder={0 0 0},pdfborderstyle={},backref=false,colorlinks=false]
 {hyperref}
\hypersetup{pdftitle={Your Title},
 pdfauthor={Your Name},
 pdfpagelayout=OneColumn, pdfnewwindow=true, pdfstartview=XYZ, plainpages=false}

\makeatletter

\floatstyle{ruled}
\newfloat{algorithm}{tbp}{loa}
\providecommand{\algorithmname}{Algorithm}
\floatname{algorithm}{\protect\algorithmname}

\usepackage[caption=false,font=footnotesize]{subfig}
\usepackage{algorithm}
\usepackage{algorithmic}

\usepackage{multirow} 
\usepackage{amsmath} 
\usepackage{xcolor}

\renewcommand{\algorithmicrequire}{\textbf{Input:}} 
\renewcommand{\algorithmicensure}{\textbf{Output:}}

\allowdisplaybreaks[4]

\ifCLASSOPTIONcompsoc
\usepackage[caption=false,font=normalsize,labelfont=sf,textfont=sf]{subfig}
\else
\usepackage[caption=false,font=footnotesize]{subfig}
\fi

\usepackage{cite}
\usepackage{bm}
\usepackage{algorithmic}
\usepackage{algorithm}
\usepackage{graphicx}
\IEEEoverridecommandlockouts

\usepackage{lettrine}

\usepackage{geometry}
\@ifundefined{showcaptionsetup}{}{%
 \PassOptionsToPackage{caption=false}{subfig}}
\usepackage{subfig}
\makeatother
\begin{document}
\captionsetup[figure]{labelfont={},name={Fig.},labelsep=period}
\title{\textcolor{black}{UAV Swarm Deployment and Trajectory for 3D Area Coverage via Reinforcement Learning}}

\author{\IEEEauthorblockN{Jia He$^{	\textrm{1, 2}}$, Ziye Jia$^{\textrm{1, 2}\star}$, Chao Dong$^{\textrm{1}}$, Junyu Liu$^{\textrm{2}}$, Qihui Wu$^{\textrm{1}}$, and Jingxian Liu$^{\textrm{3}}$\\
 }\IEEEauthorblockA{$^{\textrm{1}}$The Key Laboratory of Dynamic Cognitive System of Electromagnetic Spectrum Space, Ministry of Industry and\\
Information Technology, Nanjing University of Aeronautics and Astronautics, Nanjing, Jiangsu, 210016\\
$^{\textrm{2}}$State Key Laboratory of Integrated Services Networks (Xidian University), Xi'an, 710071\\
$^{\textrm{3}}$Ericsson (China) Communications Company Ltd., Beijing, 100102 \\
$^{\star}$Email: jiaziye@nuaa.edu.cn\\}
\thanks{{This work was supported in part by the Natural Science Foundation
of Jiangsu Province of China under Project BK20220883, and  in part by the Natural Science Foundation on Frontier Leading Technology Basic Research Project of Jiangsu BK20222013.
}}

}

\maketitle

\thispagestyle{empty}
\pagestyle{empty}

\begin{abstract}
Unmanned aerial vehicles (UAVs) are recognized as promising technologies for area coverage due to the flexibility and adaptability. However,  the ability of a single UAV is limited, and as for the large-scale three-dimensional (3D) scenario, UAV swarms can establish seamless wireless communication services. Hence, in this work, we consider a scenario of UAV swarm deployment and trajectory to satisfy 3D coverage considering the effects of obstacles. In detail, we propose a hierarchical swarm framework to efficiently serve the large-area users. Then, the problem is formulated to minimize the total trajectory loss of the UAV swarm. However, the problem is intractable due to the non-convex property, and we decompose it into smaller issues of  users clustering, UAV swarm hovering points selection, and swarm trajectory determination. Moreover, we design a Q-learning based algorithm to accelerate the solution efficiency. Finally, we conduct extensive simulations to verify the proposed mechanisms, and the designed algorithm outperforms other referred methods.

\end{abstract}


\section{Introduction}

\lettrine[lines=2]{U}{nmanned} aerial vehicles (UAVs) are recognized as promising solutions to tackle area coverage problems (ACPs) due to the inherent mobility and flexibility.
In addition, UAVs are widely used in various practical scenarios, such as wireless data collection, mapping, disaster rescue, reconnaissance and surveillance \cite{b1,b2,b3,b4}. However, as for a single UAV, the limitations of payload, size, energy capacity, etc., impact its service ability \cite{c6}. Thus, a single UAV may not ensure sufficient and reliable wireless communication services for ground users (GUs) \cite{c5}. Fortunately, the UAV swarm, consisting of multiple small UAVs, can establish seamless wireless communication services for large-area GUs. However, the following challenges related to the UAV swarm still should be handled: 1) the efficient cooperation of UAVs to serve multiple GUs; 2) how to deploy the UAV swarm to achieve a tradeoff between coverage and energy efficiency; 3) the trajectory planning for the swarm considering the effects 
of obstacles. Hence, it is essential to design a reasonable UAV swarm framework and efficient trajectory plan for large-scale 3D area coverage.

There exist a couple of recent works related to UAVs in ACPs scenarios. For instance, Zhao \textit{et al.} in \cite{b5} study the 3D mobile sensing and trajectory optimization problem with a single UAV.  In \cite{b6},  Zeng \textit{et al.} investigate energy-efficient UAV communication by deriving a paradigm related to UAV's energy consumption and trajectory planning. He \textit{et al.} in \cite{b7} consider the 3D UAV deployment problem in uneven terrains for GUs' connectivity and coverage. Jia \textit{et al.} in \cite{b8} present a matching game theory-based algorithm to deal with the offloading decisions from users to UAVs. Hou \textit{et al.} in \cite{b9} propose a joint neural network to optimize both the transmission power and the user association in a multi-UAV communication system with moving users. In \cite{b10}, the utilization of UAV swarms shows  improved coverage performance for ACPs. However, as far as the authors' knowledge, the above works lack considering deployment and trajectory planning for UAV swarms, which is a significant issue.

Hence, in this work, we consider a large-scale 3D coverage scenario with non-uniformly distributed GUs. Then, we design a hierarchical UAV swarm framework consisted of a cluster head UAV (H-UAV) and tail UAVs (T-UAVs) to efficiently accomplish the coverage mission. In particular, we divide the large-scale 3D area into discrete space considering obstacles. Then, the ACP is formulated as an optimization problem to minimize the total trajectory loss of the UAV swarm. Due to the non-convex property, we further decompose it into smaller problems of GUs clustering, UAV swarm hovering points (HPs) selection and trajectory planning. Then, we design a K-means clustering based method to cluster GUs and select the optimal HPs as deployment positions to satisfy the coverage requirements. Then, we model the trajectory planning problem as a Markov decision process (MDP) to depict the complex environment. Then, a Q-learning based UAV swarm trajectory planning algorithm (QLUTP) is designed to handle the problem. Finally, we conduct extensive simulations to verify the effectiveness of  the proposed methods.

This paper is organized as follows. In Section \ref{Section2}, we present the system model of large-scale 3D area coverage and provide the problem formulation. Section \ref{Section3} proposes the algorithms of deployment and trajectory planning for the UAV swarm. Further, Section \ref{Section4} conducts simulations and analyzes the results to verify the proposed mechanisms. Finally, conclusions are drawn in Section \ref{Section5}. 

\begin{figure}
\centering
	\includegraphics[width=0.48\textwidth,height=0.27\textwidth]{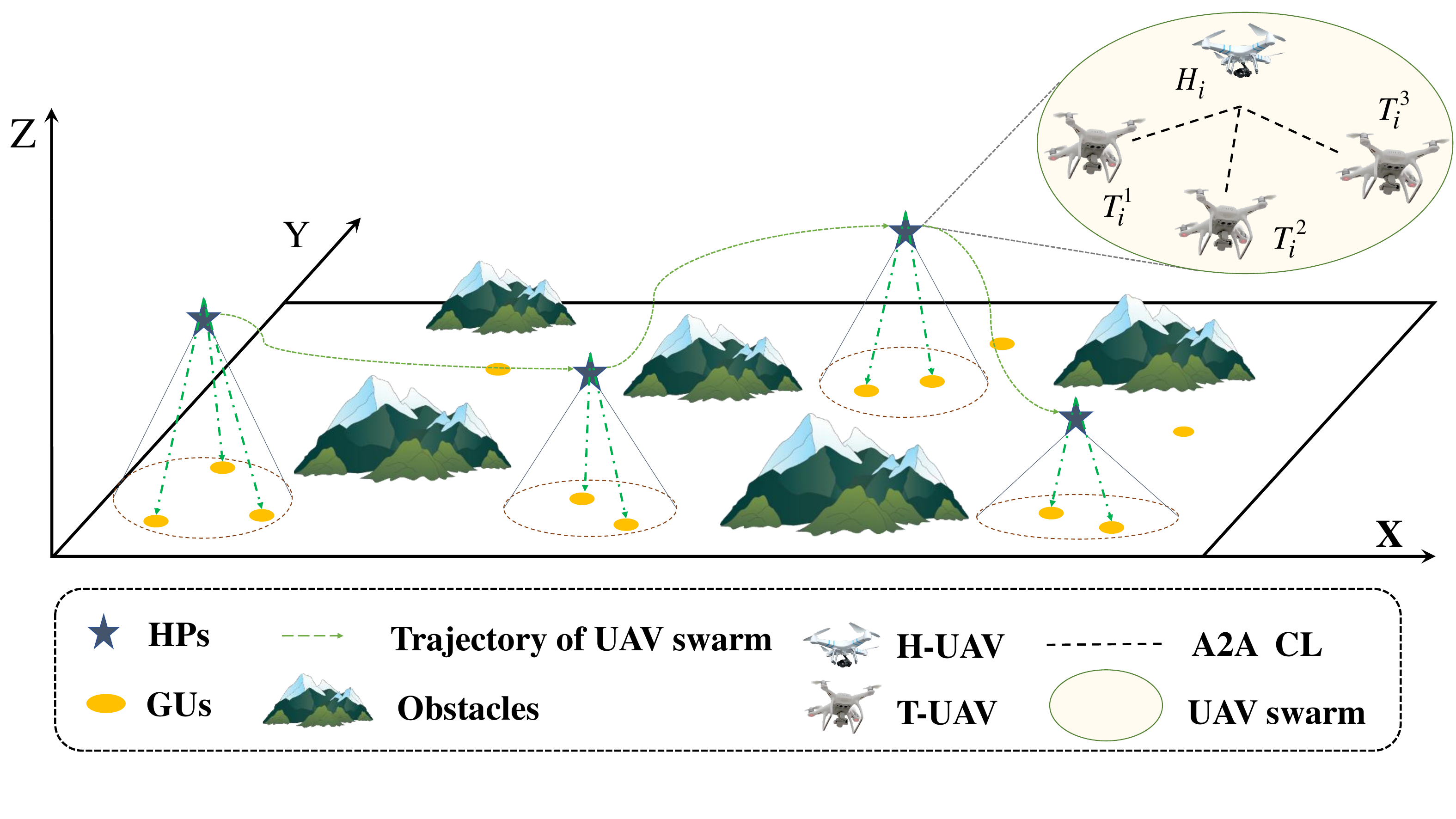}
\caption{UAV swarm enabled 3D area coverage scenario.}
\label{UAV swarm enabled 3D area coverage scenario.}

\end{figure}
\vspace{2pt}
\section{System model and problem formulation} \label{Section2}

\subsection{3D Area Coverage Scenario} 
As shown in Fig. \ref {UAV swarm enabled 3D area coverage scenario.}, we consider a UAV swarm enabled coverage scenario, including a UAV swarm and a set of GUs. In general, it is assumed that all UAVs fly with a constant speed at the same altitude $h$ within $[{h_{\min }},{h_{\max }}]$. Further, GUs are non-uniformly distributed in the coverage area, denoted as $D=({{u_1},{u_2},...,{u_M}})$.  During the flight, the UAV swarm needs to select $N$ HPs, i.e., $\Omega  = \{ {q_1},{q_2},...,{q_N}\}$, to satisfy the coverage requirements.

In this work, we adopt the regular grid method to describe the large-scale 3D space, denoted as ${\cal G}$ \cite{zzx}. Specifically, ${\cal G}$ is abstracted as an ${L_x} \times {L_y} \times {L_z}$ cuboid composed of  small cubes with length of $\Delta l$. The center coordinates of the cube are presented as $(x,y,z)$. Further, the number of cubes is ${N_x} \times {N_y} \times {N_z}$, i.e.,
\begin{equation}\label{e1}
    {N_x} = \left\lceil {\frac{{{L_x}}}{{\Delta l}}} \right\rceil ,{N_y} = \left\lceil {\frac{{{L_y}}}{{\Delta l}}} \right\rceil ,{\rm{and \ }}{N_z}{\rm{ = }}\left\lceil {\frac{{{L_z}}}{{\Delta l}}} \right\rceil ,
\end{equation}
in which the ceil function $\lceil f(x)\rceil $ returns the value of a number rounded upward to the nearest integer. The partition accuracy $\Delta l$ is a hyper-parameter, and smaller $\Delta l$ indicates that the accuracy of space ${\cal G}$ increases. However, due to the searching and storage consumption, the computational complexity generally grows exponentially with slight decline of $\Delta l$. Therefore, it is essential to make comprehensive analysis according to the performance constraints of the UAV swarm and the practical scale of ACPs.

Moreover, it is necessary to consider the distribution of obstacles in the 3D space ${\cal G}$ to avoid unnecessary collisions. In detail, the obstacle is described as $\left\{ {{\rm O}{ \rm |} ({x^o},{y^o}),Ob({x^o},{y^o})} \right\}$, where $({x^o},{y^o})$ is the projection of a certain point on the horizontal plane, and $Ob({x^o},{y^o})$ is the height of the corresponding obstacle point. Further, flag ${\cal L}_{ob}$ represents the collision, i.e., 
 \begin{equation}
{{\cal L}_{ob}} = \left\{ \begin{array}{l}
0,\quad{\rm{    }}{\cal C} \notin {\rm O},\\
1,\quad{\rm{    }}{\cal C} \in {\rm O},
\end{array} \right.
\end{equation}
where ${\cal C}$ is the position of the UAV swarm. ${\cal L}_{ob}=0$  means there exist no collisions between UAVs and obstacles. Otherwise, ${{\cal L}_{ob} = 1}$ represents there exist collisions.

\begin{figure}[t]
\centerline{\includegraphics[width=0.5\textwidth,height=0.28\textwidth]{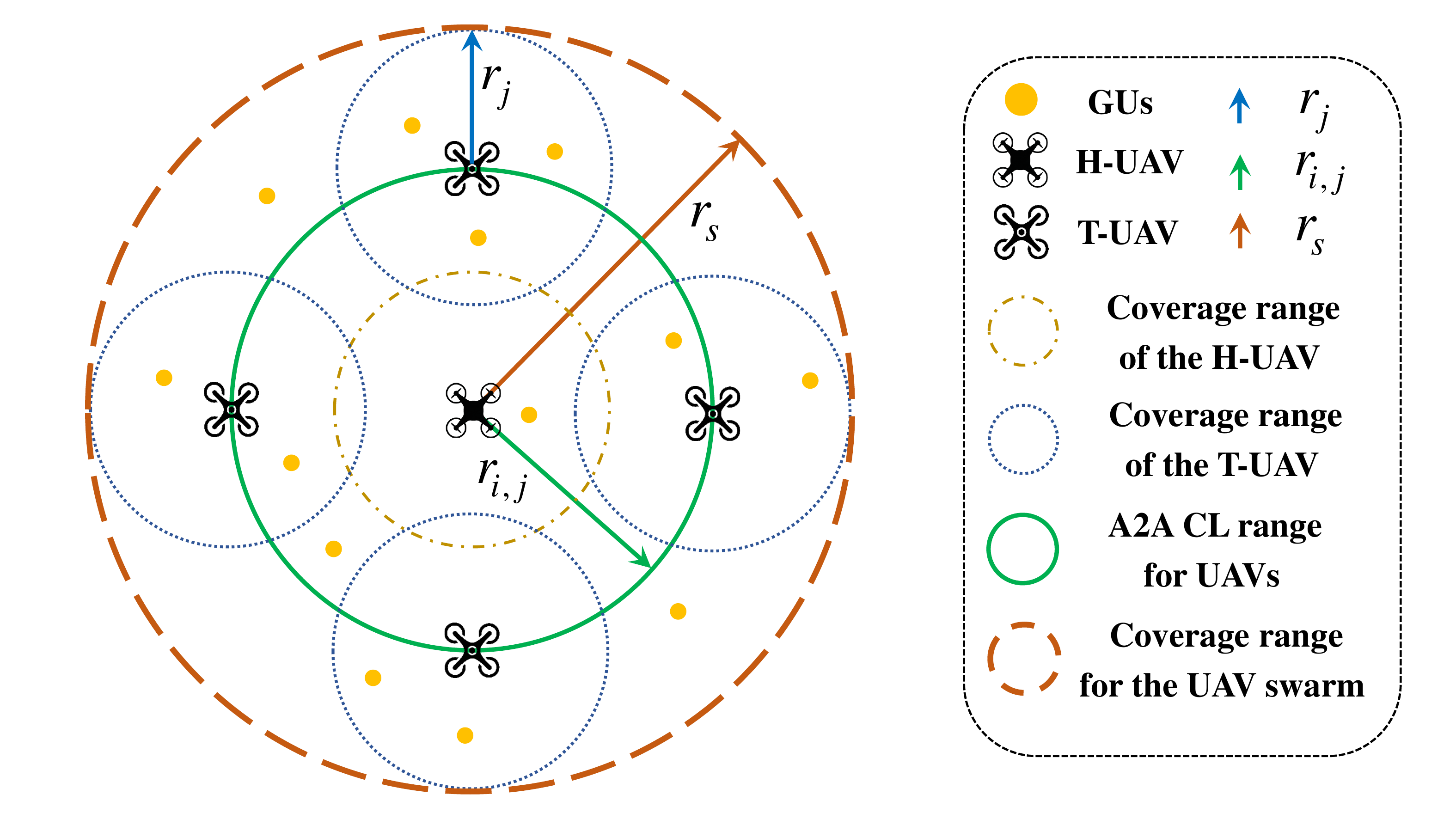}}
\caption{ An illustration of the coverage rate within a UAV swarm.}
\label{f2}
\end{figure}

Theoretically, the UAV swarm can  fly above each GU for area coverage. However, this deployment policy results in additional energy consumption due to repeated coverage. Also, there are some isolated GUs which cannot be effectively covered. Therefore, a tradeoff between the coverage and energy performance should be considered. We define the GUs coverage rate $\varUpsilon$ as an index to evaluate the coverage performance of  the UAV swarm. In detail, 

\begin{equation}\label{e3}
\varUpsilon  = \frac{1}{M}\sum\limits_{n = 1}^N {{A_n}},
\end{equation}
in which $A_n$  is the number of covered GUs at $n$th HP, i.e., ${q_n} \in \Omega $. ${M}$ represents the total number of GUs located in the area.  In Fig. \ref{f2}, an illustration of the coverage rate within a UAV swarm is provided. Since the coverage radius ${r_s}$ of the UAV swarm is related with $\varUpsilon$, it is calculated as:

 \begin{equation}\label{e4}
{r_s} = {r_j} + {r_{i,j}},
\end{equation}
where $r_{i,j}$ represents the communication radius between H-UAV and T-UAV, and $r_j$ is the coverage radius of single T-UAV.

\subsection{ UAV Swarm Model}\label{AA}

We propose a hierarchical UAV swarm framework composed of multiple small low-cost UAVs. All UAVs can be divided into two categories according to different roles in the swarm.
 H-UAV, acting as the leader of the swarm, can establish wireless communication links with other UAVs or ground base station for data  transmission. T-UAVs, acting as followers, perform the coverage mission with the leadership of H-UAV. Thus, the UAV swarm is  defined as $ \{H_{i},T_{i,1},...,T_{i,j},...,T_{i,K}\} \label{eq}$, in which ${H_i}$ and  ${T_{i,j}}$ represent H-UAV and $j$th T-UAV of swarm $i$, respectively. In the swarm, we adopt the star topology, i.e., the H-UAV serves as the communication center and directly communicates with all T-UAVs. Besides, it is assumed that T-UAVs can only communicate with the H-UAV within the same swarm.

\subsection{Communication Model}

\textit{1)\enspace A2A Communication Model:}
We adopt the typical air-to-air (A2A) communication link (CL) model to characterize the communication model between the H-UAV and T-UAV \cite{011}. Firstly, the position of H-UAV ${H_i}$ and T-UAV ${T_{i,j}}$ are denoted by ${{\cal I}_i}$ and ${{\cal I}_{i,j}}$, respectively.
Then, the received power of H-UAV from T-UAVs is:
\begin{equation}\label{e5}
  {P_{r,H}}\left( {{{\cal I}_i},{\rm{ }}{{\cal I}_{i,j}}} \right) = {P_{C,T}} + {G_t} + {G_r} - {P_l}\left( {{{\cal I}_i},{\rm{ }}{{\cal I}_{i,j}}} \right),
\end{equation}
where ${G_t}$ and ${G_r}$ are the constant antenna gains of T-UAV and H-UAV, respectively. ${P_{C,T}}$ is the constant transmission power for UAVs. 
${P_l}\left( {{{\cal I}_i},{\rm{ }}{{\cal I}_{i,j}}} \right)$ is the large-scale fading coefficient, defined as:
\begin{equation}\label{e6}
{P_l}\left( {{{\cal I}_i},{\rm{ }}{{\cal I}_{i,j}}} \right) = 10 a {\log _{10}}\left( {\frac{{4\pi {{\left\| {{{\cal I}_i} - {{\cal I}_{i,j}}} \right\|}_2}{f_c}}}{{{v_c}}}} \right),
\end{equation}
 where $a > 0$ is the trajectory loss exponent, ${v_c}$ is the speed of light, and ${f_c}$ is the electromagnetic wave frequency \cite {012}.

In addition, the H-UAV and T-UAV establish a CL if the received power satisfies: 
\begin{equation}\label{e7}
{P_{r,H}}\left( {{{\cal I}_i},{\rm{ }}{{\cal I}_{i,j}}} \right) \ge {P_0},
\end{equation}
where ${P_0}$ is set as the threshold power. Based on (\ref{e7}), the Euclidean distance ${\left\| {{{\cal I}_i} - {{\cal I}_{i,j}}} \right\|_2}$ between ${H_i}$ and ${T_{i,j}}$  satisfies:
\begin{equation}\label{e8}
{\left\| {{{\cal I}_i} - {{\cal I}_{i,j}}} \right\|_2} \le \frac{{{v_c}}}{{4\pi {f_c}}}\exp \left\{ {\frac{{\ln (10)}}{{10a}}\left( {{P_{C,T}} + {G_t} + {G_r} - {P_0}} \right)} \right\}.
\end{equation}
Note that the horizontal distance between ${H_i}$ and ${T_{i,j}}$ is equal to the Euclidean distance,  ${r_{i,j}} = {\left\| {{{\cal I}_i} - {{\cal I}_{i,j}}} \right\|_2}$.

\textit{2)\enspace A2G Link Model:}
As shown in Fig. \ref{UAV swarm enabled 3D area coverage scenario.}, T-UAVs provide effective signal coverage for corresponding GUs in the target area. In the air-to-ground (A2G) communication model, the channel gain of the line-of-sight link between T-UAVs and GUs is: 
\begin{equation}\label{e9}
h_{j,m}^{A2{\rm{G}}} = \frac{{{\beta _0}}}{{d_{j,m}^2}},
\end{equation}
where ${\beta _0}$ is the channel gain at the reference distance of each meter, and $d_{j,m}$ represents the Euclidean distance between ${T_{i,j}}$ and $u_m$ \cite{013}. Let ${P_{m,j}}$ represent the transmitting power of GU $u_m$, and the transmission rate is expressed as: 
\begin{equation}\label{e10}
{R_{j,m}} = B\log 2\left( {1 + \frac{{{P_{m,j}}h_{j,m}^{A2{\rm{G}}}}}{{B\eta }}} \right),
\end{equation}
where $B$ and $ \eta $ are channel bandwidth and noise power density, respectively.
 The transmission rate ${R_{j,m}}$ should be not less than the minimum achievable rate threshold ${R_{Th}}$, i.e.,
 \begin{equation}\label{e11}
{R_{j,m}} \ge {R_{Th}},
\end {equation}
 which guarantees that the T-UAV can establish a stable CL with the GU. According to (\ref{e9}) and (\ref{e10}), we obtain the maximum coverage range  $d_{j,m}$ of T-UAV as: 
\begin{equation}\label{e12}
{d_{j,m}}{\rm{ }} = {\rm{ }}\sqrt {{{{P_{m,j}}{\beta _0}} \mathord{\left/
 {\vphantom {{{P_{m,j}}{\beta _0}} {({2^{{{{R_{j,m}}} \mathord{\left/
 {\vphantom {{{R_{j,m}}} B}} \right.
 \kern-\nulldelimiterspace} B}{\rm{  }} - 1}}}}} \right.
 \kern-\nulldelimiterspace} {({2^{{{{R_{j,m}}} \mathord{\left/
 {\vphantom {{{R_{j,m}}} B}} \right.
 \kern-\nulldelimiterspace} B}{\rm{  }} - 1}}}}{\rm{) }}},
\end{equation}
where the transmission rate ${R_{j,m}}$ is equal to threshold ${R_{Th}}$. Then, the coverage radius of a single T-UAV is calculated as:
 \begin{equation}\label{e13}
{r_j} = \sqrt {{d_{j,m}}^2 - {h}^2},
\end{equation}
where $h$ is the flight altitude of the UAV swarm.

\subsection{Problem Formulation}

Since ACP aims to save energy via appropriate deployment of the UAV swarm, the objective is to minimize the flight trajectory loss of the UAV swarm. Thus, the optimization problem is detailed as:
\begin{subequations}
\begin{align}\label{e14}
{ \textbf{\textbf{P0}}}:\quad\mathop {\min }\limits_\rho  \quad&{\rm{     ( }}\Delta l\sum\limits_{\rho  \in {{\cal P} }} {Ed_n^{n + 1}} {)_{\Omega ({q_1} \to {q_N},\varUpsilon )}},\\
{\rm{ s}}{\rm{.t.}}\quad&\text{C1}{\rm{ : }}\quad\rho  \in {\cal P},\\
&\text{C2}{\rm{ : }}\quad{{\cal L}_{ob}} = 0,\\
&\text{C3}-\text{C4}{\rm{ : }}\quad (\ref{e7}),\ (\ref{e11}),\\
&\text{C5}{\rm{ : }}\quad{\varUpsilon } \ge \varUpsilon _{Th},
\end{align}
\end{subequations}
where variable $\rho $ is a swarm trajectory path, $\Delta l$ is trajectory loss for each step, and $Ed_n^{n + 1}$ is the number of path steps from $q_n$ to $q_{n+1}$. All the HPs satisfying the coverage rate ${\varUpsilon}$ are denoted as $\Omega ({q_1} {\to} {q_N},{\rm{ }}{\varUpsilon ^{{\rm{ }}}})$. Further, in constraint C1, the UAV swarm should fly within the tolerable space, i.e., $\rho  \in {\cal P}$. Constraint C2 addresses the distribution of obstacles, in which flag ${\cal L}_{ob}=0$ means there exist no collisions. As illustrated in C3, the received power ${P_{r,H}}$ should be not less than the limited threshold $P_0$. In constraint C4, transmission rate $R_{j,m}$ should not less  than threshold $R_{Th}$. Constraint C5 guarantees that the selected HPs satisfy the limitation of coverage rate ${\varUpsilon }$. Thus, ${\varUpsilon }_{Th}$ is set to ensure the basic coverage performance.
Due to the objective is to minimize the flight trajectory loss of the UAV swarm, both the deployment positions selection and trajectory planning are non-convex. As for the problem complexity, the solution space grows exponentially with the problem scale due to the non-convex property of \textbf{P0}, which is intractable via a traditional optimization approach.

\section{ Deployment and Trajectory Planning for the UAV swarm}\label{Section3}

To efficiently deal with \textbf{P0}, we decompose it into smaller problems of GUs clustering,  HPs selection for UAV swarm and trajectory planning. Firstly, in Section \ref{section3a}, a K-means based method is designed to cluster GUs, and select multiple HPs as deployment positions for the UAV swarm.
Then, the ACP is modeled as an MDP in Section \ref{section3b}. Further, a Q-learning based UAV swarm trajectory planning  (QLUTP) is designed in Section \ref{section3c} to minimize the flight trajectory loss.

\subsection{GUs Clustering and HPs Selection}\label{section3a}

We adopt the K-means clustering based method to select optimal deployment positions of HPs. The variable $\Omega ({q_1}{\to}{q_N},\varUpsilon )$ and $N$ depend on the dispersion degree of GUs and the scale of the coverage area. The details of GUs clustering and HPs selection approach are presented in Algorithm \ref{a1}. Firstly, we initialize the deployment of GUs $\left\{ {{u_1},{u_2},...,{u_M}} \right\}$, the coverage radius ${r_s}$ and number of clusters $N$.
From step 3 to 12, GUs are clustered according to the distance and obtain the updated clusters. Further, we analyze the coverage range conditions of GUs in each cluster, as shown in step 13. Finally, we obtain the deployment positions of HPs satisfying coverage constraints, as illustrated from step 14 to 15.

\begin{algorithm}[!t]
    \caption{GUs Clustering and Hovering Points Selection } 
    \begin{algorithmic}[1]\label{a1}
        \STATE \textbf{Initialize} the deployment of GUs $\left\{ {{u_1},{u_2},...,{u_M}} \right\}$, the coverage radius ${r_s}$ and the number of clusters $N$.
        \STATE Create $N$ points randomly as the starting centroid.
        \FOR {$episode = 1$ to ${MAX}_{eps}$}
        \STATE  Reset the starting clusters $\left\{{C_1,C_2,...,C_N}\right\}$ as $\emptyset$.
		\FOR {GUs {$i \in \left[1,M\right]$} }
		\STATE Calculate the Euclidean distance between the centroid and each data point.
        \STATE Assign the data point to the nearest cluster.
        \ENDFOR
        \FOR {cluster $n \in \left[ 1,N \right]$ }
        \STATE Reset the centroid of each cluster.
        \ENDFOR
           		\ENDFOR
\STATE Count the number of data points within the coverage radius ${r_s}$ for each cluster.
\STATE Calculate the coverage rate $\varUpsilon$ according to (\ref{e3}). 
\STATE \textbf{Return} the positions of updated clusters $\left\{{C_1,C_2,...,C_N}\right\}$ and the coverage rate $\varUpsilon$.

    \end{algorithmic}
\end{algorithm}

\vspace{6pt}

\subsection{MDP for Trajectory Planning}\label{section3b}

Reinforcement learning (RL) studies the sequential decision-making process, in which the agent acts as the subject body and interacts with the objective environment. Correspondingly, the UAV swarm is abstracted as an agent in the environment $\cal G$ with obstacles. Thus, the optimization problem \textbf{P0} is reformulated in the form of MDP. 
An MDP problem is defined by a tuple $\left\{ {S,A,P,R,\gamma } \right\}$, where $S$ is the state space. $A$ is the action space. $\left\{{P:S \times A \to \Delta (S)}\right\}$ is the state transition probability function from state $s\in S$ to another state $s'\in S$ with action $a \in A$.  $\left\{ {R:S \times A ={\cal R} } \right\}$ is the immediate reward function of the agent, and $\gamma  \in [0,1]$ is a discounted factor. In addition, $\pi $ represents the policy, and ${Q_\pi }(s,a)$  represents the action value function.

\begin{algorithm}[!t]
    \caption{ QLUTP for Trajectory Planning  }
    \label{a2}
    \renewcommand{\algorithmicrequire}{\textbf{Input:}}
    \renewcommand{\algorithmicensure}{\textbf{Output:}}
    \begin{algorithmic}[1]
        \REQUIRE $N$,$\enspace\alpha ,\enspace\gamma,\enspace\varepsilon {\rm{\enspace and   }}\enspace\Delta l$.
        \ENSURE Optimal policy ${\pi ^*}$.
        \STATE \textbf{Initialize} action value table  $\cal Q$ arbitraryly according to ${ Q}(s,a),\forall s \in { S},a \in {A}(s){}{\rm{ }}$.
        \STATE \textbf{Repeat} (for each episode)
        \STATE \quad Initialize  state space  ${S}$.
        \begin{ALC@g}
            \STATE Choose action ${a \in A}$ under ${s \in S}$ with the proposed policy ${\rm{\varepsilon(ep){-}greedy}} $ derived from table ${\cal Q}$.	
        \end{ALC@g}	
        \STATE \quad Take action ${A}$, observe ${ R}$ and ${ S^{'}}$. 
        \begin{ALC@g}
            \STATE  Update table ${\cal Q}$ according to the formulation \quad ${ Q}({ S},{ A}){\leftarrow}{ Q}({S},{ A}) {+} \alpha [{ R}{+}\gamma {\max _a}{ Q}({ S^{'}},{ A}){-}{\ Q}({ S},{ A})]$.
        \end{ALC@g}	
        \STATE  \quad ${ S} \leftarrow { S^{'}}$.
        \STATE  \textbf{Until} ${\ S}$ is terminal.

    \end{algorithmic}
\end{algorithm}

\textit{1)\enspace  State space:} The state space $S$ indicates the environment of the agent. In order to make the agent aware of the environment state, we introduce the positions of HPs $\Omega ({q_1} {\to} {q_N},{\rm{ }}{\varUpsilon ^{{\rm{ }}}})$ and obstacles $\left\{ {{\rm O}{ \rm |} ({x^o},{y^o}),Ob({x^o},{y^o})} \right\}$.

\textit{2)\enspace  Action space:} The action space provides six actions for the agent, i.e., $A = \{ {a_f},{a_b},{a_r},{a_l},{a_u},{a_d}\} $. Note that the elements represent the action of flying forward, backward, left, right, up and down, respectively. At each time step, agent selects one action from $A$.

\textit{3)\enspace  Transition probability:} The transition probability function $P$ represents the dynamic characteristic of the environment. It is intractable to model accurately due to the model-free property. 

\textit{4)\enspace  Discounted factor $\gamma $:} The discounted factor $\gamma $ denotes the effect of the future reward, and a larger $\gamma $ indicates the decision of focusing on long-term reward.

\textit{5)\enspace  Reward function:} The reward function aims to lead the agent for achieving the goals of the task in RL process. 
Further, ${R_{t + 1}}$ is denoted as instantaneous reward of performing action ${a_t}$ under current state ${s_t}$, i.e.,
\begin{equation}
{R_{t + 1}}({s_t},{a_t}) = ( - 0.1 \times \Delta l)(1 - {{\cal H}_n}) + 100{{\cal H}_n} - 100{{\cal L}_{ob}},
\end{equation}
where $\Delta l$ is the trajectory loss for each step. ${{\cal H}_n}$ is defined as the flag for the end of the trajectory planning process, i.e.,
 \begin{equation}
{{\cal H}_n} = \left\{ \begin{array}{l}
0,\quad{\rm{    }}{p_i} \in \Omega, \\
1,\quad{\rm{    }}{p_i} \notin \Omega.
\end{array} \right.
\end{equation}

\textit{6)\enspace Action value function:} The action value function ${Q_\pi }(s,a)$ indicates the expected future reward for performing action ${a_t}$  under the current state ${s_t}$, i.e.,
\begin{equation}
{Q_\pi }(s,a) = {E_\pi }[{R_{t + 1}} + \gamma {R_{t + 2}} + {\gamma ^2}{R_{t + 3}} + ...\left| {{s_t} = s,{a_t} = a} \right|],
\end{equation}
where ${a_t}$ and ${s_t}$ represent the current action and state, respectively. The cumulative discounted reward ${G_t}$ is denoted as:
 \begin{equation}
{G_t} = {R_{t + 1}} + \gamma {R_{t + 2}} + {\gamma ^2}{R_{t + 3}} + ...,
\end{equation}
where the reward of the future time step is implied in the present step, but future rewards should be multiplied with the discounted factor $\gamma $.

\vspace{4pt}

\subsection{ Q-learning based QLUTP }\label{section3c}
Q-learning algorithm is a fundamental RL method. In detail, the agent chooses action $a {\in} A(s)$ under state ${s}$ with the instruction of action value function ${Q }(s,a)$.  Moreover, all the action value functions ${Q_\pi }(s,a)$ are stored in the action value table ${\cal Q}$, which is updated during the training process. Then, according to the updated ${\cal Q}$, we obtain the optimal policy ${\pi ^*}$.  The goal is to maximize the accumulated reward $\cal R$, i.e.,
\begin{equation}
\begin{array}{l}
{\pi ^*} = \arg {\rm{ }}\mathop {\max }\limits_\pi  {\rm{ }} \cal R.
\end{array}
\end{equation}
According to the principle of value based Q-learning algorithm, policy ${\pi ^*}$ depends on the action value table ${\cal Q}$. Thus, the policy ${\pi ^*(a|s)}$ leads the agent to perform the action $a'$ under state $s'$ with the maximum ${Q}(s',a')$, i.e.,
\begin{equation}
{\pi ^*}(a|s) = \left\{ \begin{array}{l}
1,\quad a = \mathop {\arg \max _{a' \in A(s')}}Q\left( {s',a'} \right), \\
0,\quad \text{otherwise},
\end{array} \right.
\end{equation}
where  ${\max _{a' \in A(s')}}Q\left( {s',a'} \right)$ represents the max value when selecting action  $ a'$ at $s'$.

The details of the updating process of action value table ${\cal Q}$ are presented in Algorithm \ref{a2}. In detail, as in step 6, it is necessary to update $Q\left( {{s_t},{a_t}} \right)$ at time step $t$ in the learning process, according to
\begin{equation}
Q\left( {{s_t},{a_t}} \right) \leftarrow Q\left( {{s_t},{a_t}} \right) + \alpha {\delta _t},
\end{equation}
where $\alpha$ is the learning rate, and the action value error ${\delta _t}$ is denoted as 
\begin{equation}
{\delta _t} = {R_{t + 1}} + \gamma {\max _{a' \in A(s')}}Q\left( {s',a'} \right) - Q\left( {{s_t},{a_t}} \right).
\end{equation}

\begin{figure}[t]
    \centerline{\includegraphics[width=0.54\textwidth,height=0.42\textwidth]{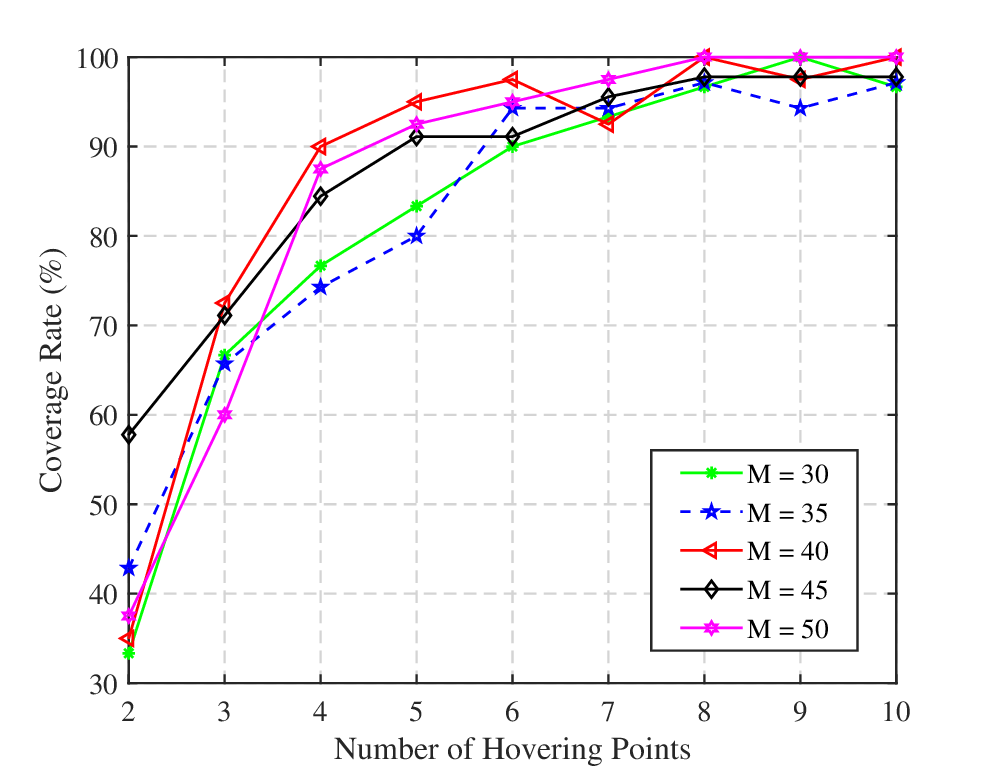}}
    \caption{ Performance of coverage with different number of hovering points (Coverage rate $v.s.$ Number of hovering points).}
    \label{Simulation results of HPs selection.}
    \end{figure}

The agent performs the action $a'$ corresponding to the optimal value ${\max _{a' \in A(s')}}Q\left( {s',a'} \right)$ under the next state $s'$ when updating ${Q}(s,a)$, resulting in an overestimation problem concerning the sampled action. Therefore, action selection strategy $\varepsilon{-}{\rm{greedy}}$ is introduced in traditional Q-learning method, i.e.,
\begin{equation}
{\pi ^*}(a|s) = \left\{ {\begin{array}{*{20}{l}}
{1 - \varepsilon ,\quad a = \mathop {\arg \max _{a' \in A(s')}}Q\left( {s',a'} \right) },\\
{\varepsilon  ,\quad {\rm{     }}a = {\rm{ random \ action}}},
\end{array}} \right.
\end{equation}
where $\varepsilon {\in} (0,1)$ represents the probability of exploring random action $a$. 

However, the traditional Q-learning method performs not well with respect to the convergence, due to the slow updating speed of ${\cal Q}$. In order to accelerate the update speed of $\cal Q$, an improved action selection strategy $\varepsilon(ep){-}{\rm{greedy}}$ is introduced. Further, the probability of random selection $\varepsilon (ep)$ is a function related to episodes, instead of the constant value. With the increment of episodes, the probability of selecting random action decreases, which results in faster updating of ${\cal Q}$. Thus, we design the QLUTP method to obtain an optimal trajectory $\rho $ to minimize the total flight trajectory loss during the coverage missions.

\begin{figure}[t]
    \centerline{\includegraphics[width=0.54\textwidth,height=0.42\textwidth]{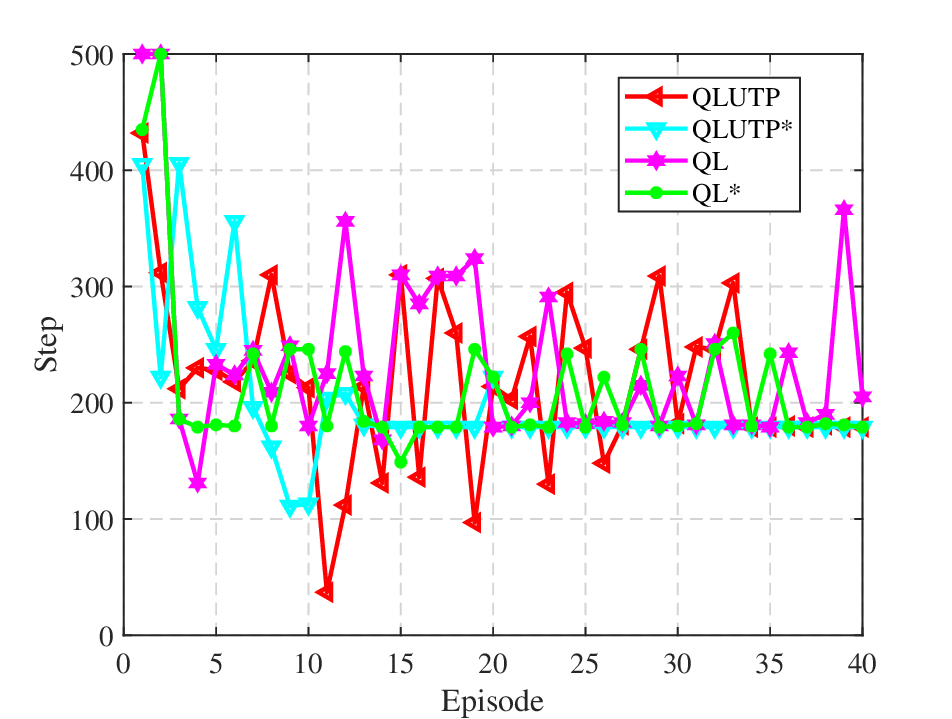}}
    \caption{Performance of convergence with different methods (Step $v.s.$ Episode).}
    \label{Step $v.s.$  Episode.}
    \end{figure}
    
    \begin{figure}[t]
    \centerline{\includegraphics[width=0.54\textwidth,height=0.46\textwidth]{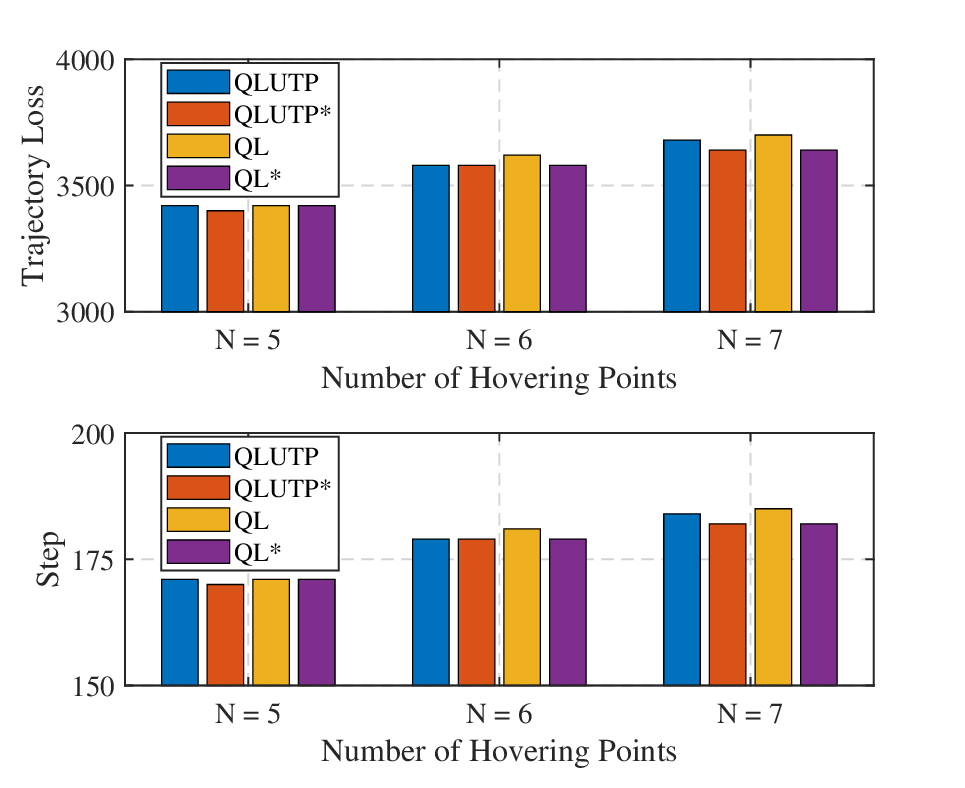}}
    \caption{ Performance of trajectory planning with different methods.}
    \label{ Trajectory planning performance.}
    \end{figure}

\vspace{4pt}

\section{Simulation results}\label{Section4}

The performance of the designed algorithms are evaluated in this section. Note that the locations of GUs are usually abstracted to a specific point process to capture and depict the positions in the coverage scenario. The poisson point process (PPP) is widely
employed to model the distribution of GUs. Therefore, the simulations are conducted by randomly deploying the GUs following PPP. In particular, we consider a 3D coverage space (2,000m${\times}$2,000m${\times}$200m), and the partition accuracy is set as $\Delta l = 20$m. The flight altitude ranges from 120m to 180m. In addition, the coverage radius of the UAV swarm is 500m. Further, the corresponding values of the simulation parameters about QLUTP are initially set as ${(\alpha ,\gamma ,episode)}={(0.6,0.6,40)}$. In addition, MATLAB is used as the platform to evaluate the convergence and effectiveness of the proposed mechanisms.

As illustrated in Fig. \ref{Simulation results of HPs selection.}, the coverage rate rises obviously with the increment of the number of hovering points $N$, and it gradually converges and  oscillates within a certain range slightly. The coverage rates under several scenarios with different number of GUs, i.e., ($M = [30,35,40,45,50]$), show that the average coverage rate is evidently above 90\% when $N=6$. Finally, the optimal hovering points ${\Omega {({q_1} {\to} {q_6},\varUpsilon )}}$ is obtained.

To evaluate the performance of the proposed trajectory planning method, Fig. \ref{Step $v.s.$  Episode.} indicates the convergences of the proposed QLUTP and QLUTP* obtained by different action selection strategies $\varepsilon{*}(ep)$, compared with other methods. Note that the fastest convergence of QLUTP* reveals the superiority and effectiveness of Algorithm \ref{a2}.  Moreover, the trajectory loss of different algorithms for three scenarios are shown in Fig. \ref{ Trajectory planning performance.}. Specifically, we select four methods for trajectory planning under different HPs sets. It is shown that the distance of trajectory increases significantly with $N$ increasing, which indicates that the UAV swarm consumes more flight energy. In addition, the optimal trajectory planned by QLUTP* needs fewer steps with less trajectory loss.

\section{Conclusions}\label{Section5}

In this work, we propose a hierarchical framework for the UAV swarm to handle the ACP for large-scale 3D coverage scenario. The UAV swarm provides steady wireless communication for non-uniformly distributed GUs. Additionally,  K-means method is adopted  to cluster GUs and select appropriate HPs. Further, according to the MDP model, we propose a QLUTP approach for trajectory planning.  Simulations are conducted to evaluate the performance of the proposed method, and the results demonstrate that the proposed algorithms perform better for ACP than other baseline methods.

\vspace{6pt}

\bibliographystyle{IEEEtran}
\bibliography{ref}
\end{document}